\documentclass[%
superscriptaddress,
showpacs,
amsmath,amssymb,
aps,
prc,
showkeys,
twocolumn,
floatfix
]{revtex4-1}
\usepackage{amssymb}

\usepackage{amsmath,mathrsfs,amssymb}
\usepackage{eurosym}
\usepackage{graphicx,subfigure}
\usepackage{color}
\usepackage{indentfirst}
\usepackage{extarrows}
\usepackage{hyperref}
\hypersetup{colorlinks=true, citecolor=blue, urlcolor=blue, linkcolor=blue}
\usepackage{ulem}

\begin{document}

\title{Status on $\mathbf{{}^{12}{\rm C}+{}^{12}{\rm C}}$ fusion at deep subbarrier energies: impact of resonances on astrophysical  $S^{*}$ factors}

\author{C. Beck$^a$, A.M. Mukhamedzanov$^b$ and X. Tang$^c$}
\address{
$^a$D\'epartement de Recherches Subatomiques, Institut Pluridisciplinaire 
Hubert Curien, IN$_{2}$P$_{3}$-CNRS and Universit\'e de Strasbourg - 23, rue 
du Loess BP 28, F-67037 Strasbourg Cedex 2, France\\
E-mail: christian.beck@iphc.cnrs.fr\\
$^b$Cyclotron Institute, Texas A$\&$M University, College Station, Texas, 77843, USA \\
E-mail: akram@comp.tamu.edu\\
$^{c}$ Institute of Modern Physics, CAS, Lanzhou, P.R.China
Joint Department for Nuclear Physics, Lanzhou University and Institute of Modern
Physics, CAS, China \\
E-mail: xtang@impcas.ac.cn}

\begin{abstract}
Since the discovery of molecular resonances in $^{12}$C+$^{12}$C 
in the early sixties a great deal of research work has been undertaken to study 
$\alpha$-clustering and resonant effects of the fusion process at sub-Coulomb barrier
energies.  The modified astrophysical $S^{*}$ factors of $^{12}$C+$^{12}$C  fusion have been extracted from direct fusion measurements at deep 
sub-barrier energies near the Gamow window. They were also obtained by the indirect Trojan horse method (THM). A comparison of direct
measurements and the THM, which elucidates problems in the analysis of the THM, is discussed in this Letter to the Editor.
\end{abstract}

\maketitle

\section{Introduction}
In the last decades, one of the greatest challenges in nuclear science is the understanding of the clustered structure of nuclei from both experimental  and theoretical perspectives.
The role of cluster configurations in stellar He burning is well established. One of the most exciting topics of contemporary nuclear astrophysics is  the nature and the role of resonance structures that characterize the low-energy cross section of the $^{12}$C+$^{12}$C fusion process which plays a very important role in a wide variety of stellar burning scenarios such as massive stars, type Ia supernovae and superbursts. One of the possible scenarios of formation of Supernovae Ia is merging of the binary system  of two white dwarfs \cite{Mori}. The outcome of this merging scenario  is controlled by the $^{12}$C+$^{12}$C fusion. 

An extensive scientific discussion about this reaction is underway in experimental investigations  using direct \cite{Becker,Kettner,Aguilera,Spillane,Patterson,Jiang18}  and indirect THM \cite{THM(2018)}  measurements. The standard $^{12}$C+$^{12}$C reaction rate was established by Caughan and Fowler \cite{CF}. Their simple extrapolation agrees reasonably with some recent theoretical calculations, such as CC-M3Y+Rep \cite{Esbensen}, TDWP \cite{Wiescher} and barrier penetration model based on the global S\'ao Paulo potentials (SPP) \cite{SPP}, see Fig. \ref{fig_Tang}, as well with sophisticated coupled-channels calculations \cite{Chen,Khoa} and recent Hartree-Fock and time-dependent Hartree-Fock calculations \cite{Godbey}. We note that the predictions based on the CC-M3Y+Rep method are considered to be the upper limit of the $^{12}$C+$^{12}$C fusion cross section \cite{Notani,Jiang13}.

Guided by the experimental astrophysical $S$-factors
of the medium-heavy systems,  a phenomenological hindrance model was developed in \cite{Jiang},  which predicts that  the $^{12}$C+$^{12}$C modified $S^{*}$-factor (its definition is given in Section II) reaches its maximum as energy decreases and then rapidly drops becoming by many orders of magnitude smaller than the standard rates used for astrophysical modeling. However, a new measurement in a different system of  $^{12}$C+$^{13}$C confirms the trends predicted by the TDHF, CC-M3Y+Rep and disapproves the prediction by the hindrance model \cite{Zhang2019}.
 
Recently the Trojan Horse Method (THM) has been applied to determine
the $^{12}$C+$^{12}$C fusion $S^{*}$-factor at energies below $E=2.7$ MeV 
\cite{THM(2018)}.  The THM $S^{*}$-factor shows a profound rise  in both $\alpha-$ and $p-$ channels as energy $E$ decreases, which is much faster than any data and models presented in Fig. \ref{fig_Tang}.  

In this Letter to the Editor, we compare direct and indirect $S^{*}$ factors for the $^{12}$C+$^{12}$C fusion. 
We elucidate three main problems regarding the THM results presented in Ref. \cite{THM(2018)} are: the THM $S^{*}$ factor does not agree with the direct measurement\cite{Spillane}; the flat THM $S^{*}$ factor at $E> 2.2$ MeV is dominantly contributed by the non-THM mechanisms; the steep rise of the THM $S^{*}$ factor as energy decreases is the result of using the plane-wave approach in which Coulomb+nuclear interactions are neglected.

\section{Comparison of direct and indirect data}

The resonant structures at very low energies have still been identified as molecular $^{12}$C+$^{12}$C configurations in the $^{24}$Mg compound nucleus \cite{Spillane}. However, the reaction rate of this reaction is calculated using an average cross section integrated over the molecular resonance components. 

\begin{figure}[htbp]
\includegraphics[width=\columnwidth]{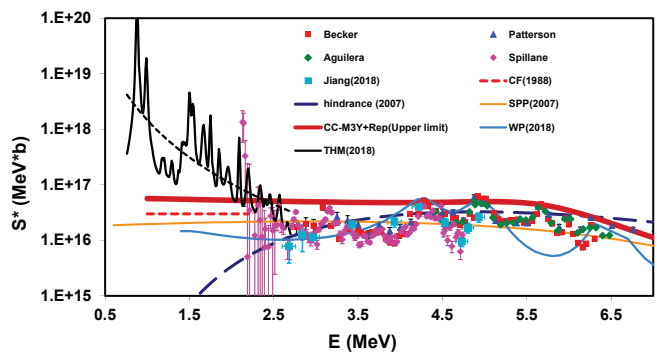}
\caption{(Color online) Total $S^{*}$ factors of the ${}^{12}{\rm C}+{}^{12}{\rm C}$ fusion. The ${}^{12}{\rm C}+{}^{12}{\rm C}$ data are from  \cite{Becker,Kettner,Aguilera,Spillane,Patterson,Jiang18}  shown as red rectangle, green dots, magenta diamonds, blue triangles and brown stars, respectively. Model calculations, CC-M3Y+Rep (thick dark red) \cite{Esbensen}, TDWP (light blue) \cite{Wiescher}, SPP (oragen) \cite{SPP}  and hindrance (blue dashed) \cite{Jiang} are also shown, respectively. The recommended averaged $S^{*}$ factor by CF88 \cite{CF} is shown as red dashed line. The THM \cite{THM(2018)} result and the fit are shown as black lines.
The uncertainty of THM $S^{*}$ factor is $\pm 21 \%$ which is not shown in the figure.
 This figure with direct, indirect data and model calculations are taken from \cite{Tang}.}
\label{fig_Tang}
\end{figure}

Fig. \ref{fig_Tang} displays the 
$S^{*}$-factors of the  $^{12}$C+$^{12}$C fusion. 
The $^{12}$C+$^{12}$C has a complicated resonance feature which continues all the way from the Coulomb barrier energy down to the lowest measured energies.These resonances have a characteristic width of about $100$ keV. Since there is no way to model the resonances in the unmeasured energy range, the averaged modified $S^{*}$-factors are used for extrapolation while the resonances are treated as fluctuations. In the literature devoted to the  $^{12}$C+$^{12}$C fusion is used  the modified astrophysical factor $\,S^{*}(E)= \sigma(E)\,E\,exp(\frac{87.21}{\sqrt{E}} +0.46\,E)$ \cite{Patterson}, where $E$ is the $^{12}$C-$^{12}$C relative energy, $\,\sigma(E)\,$ is the fusion cross-section. 

The direct measurement \cite{Spillane} reported a strong resonance at 2.14 MeV, which is the lowest achieved directly measured energy. This resonance could  be explained by a formation of  the  ${}^{12}{\rm C}+{}^{12}{\rm C}$ nuclear molecular configurations in $^{24}$Mg and the $\alpha$-cluster structure of ${}^{12}{\rm C}$. Based on the correlation between ${}^{12}$C+${}^{13}$C and ${}^{12}$C+${}^{12}$C, an upper limit is proposed for the ${}^{12}$C+${}^{12}$C $\,S^{*}$-factors\cite{Notani, Jiang, Zhang2019}. The resonance at $2.14$ MeV observed in \cite{Spillane} is 1.87$\sigma$ higher than the upper limit. 

The Argonne results \cite{Jiang18} appear to be in a qualitative agreement with  classical coupled-channel CC-M3Y+Rep calculations in \cite{Esbensen}, TDWP calculations of \cite{Wiescher}, the barrier penetration model of \cite{SPP} and more recent theoretical investigations in \cite{Chen}. Based on Fig. \ref{fig_Tang}, it is evident that the low-energy THM $S^{*}$ factor is several orders of magnitude higher than the upper limit based on the CC-M3Y-Rep calculations \cite{Zhang2019}.

\begin{figure}[htbp]
\includegraphics[width=9.0 cm]{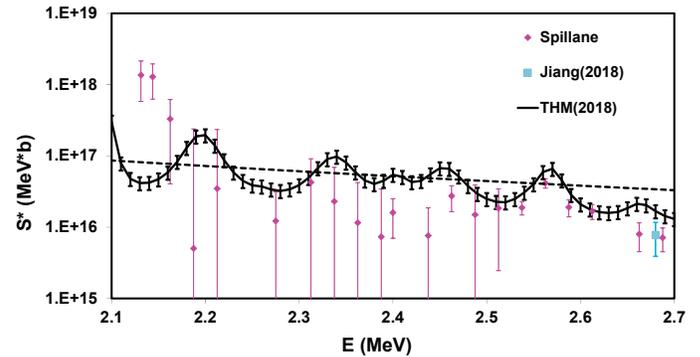}
\caption{(Color online) 
Comparison of the direct data from  \cite{Spillane} and \cite{Jiang18} with the THM data with the error bars \cite{THM(2018)} in the interval $2.1-2.7$ MeV. }
\label{fig3Tang}
\end{figure}

In Fig \ref{fig3Tang}  is shown a comparison  of the direct data from  \cite{Spillane}  and \cite{Jiang18} and indirect THM data \cite{THM(2018)}  in the energy interval $2.1-2.7$ MeV. However, such a flat oscillatory behavior of the THM $S^{*}$ factor at higher energies does not agree with the predictions obtained from the THM mechanism with the Coulomb-nuclear distortions \cite{muk2019} (see more detailed discussion in Section III).
While the direct data are more or less flat in the interval $2.16-2.67$ MeV with a possible single resonance at $2.14$ MeV, the THM $S^{*}$-factor shows an opposite trend: a rise as energy decreases from $2.67$ to $2.1$ MeV with explicit resonance structures along the entire interval. Moreover, the trend of the THM data at higher energies $E>2.1$ MeV causes the main concern because such a trend cannot be explained by the THM mechanism and can be a result of the contribution of the non-THM mechanisms (see \cite{muk2019} and discussion in Section III).

 As we know, in the THM experiment \cite{THM(2018)} only the energy dependence of the $S^{*}$ factor was measured and its absolute value was determined by the normalization of the THM data to the direct ones in the energy interval $2.50-2.63$ MeV. Then, based on the data shown in Fig. \ref{fig3Tang} we calculated the $\chi^{2}$ test of the indirect THM \cite{THM(2018)}  and direct data from \cite{Spillane,Jiang18} taking into account 19 points in the energy interval between $E=2.16$ and $2.7$ MeV. The uncertainty of the THM data is $\pm 21\%$ (see caption to Fig. \ref{fig_Tang}). The $\chi^{2}$ test is resulted in the reduced $\chi^2=4.76$. Note that for $\chi^2>2.226$, the possibility of the coincidence of data is only $0.1\%$. 
 Based on this test, we claim that the THM data disagree with the direct measurement result in the range of E=2.16-2.7 MeV.

It is worth  mentioning that the THM analysis mistakenly took the total $S^{*}$-factor of the direct measurement\cite{Spillane} as the normalization data for the $\alpha_1$ channel. As a result, the THM $S^{*}$-factors are about 30\% higher than the total $S^{*}$ obtained by the direct measurement. Even we correct this mistake, the reduced $\chi^2$ is still 2.76. 
 
One can argue that changing the normalization factor of the THM data one can move down the THM $S^{*}$ factor in the energy interval $2.16-2.7$ what would decrease the reduced $\chi^2$. We manage to drop the reduced-$\chi^{2}$ to 0.72 by normalizing the THM result by a factor of 0.4. 
But in the normalization interval  $2.50-2.63$ MeV it would result in the reduced $\chi^{2}=1.67$   increasing the discrepancy between the THM and direct data in the energy interval $2.50-2.63$ MeV and especially at $E <2.16$ MeV.  

But a more serious problem with the THM data  is that the mechanisms contributing to the THM data at $E>2.16$ MeV are not identified in \cite{THM(2018)}. 
From Figs. \ref{fig_Tang} and \ref{fig3Tang} we may conclude that both direct data and
model calculations disagree with the indirect THM data at energies below $2.2$ MeV.
 
\section{Review of  THM $S^{*}$ analysis used to extract $S^{*} $ factors for  ${}^{12}{\rm C}+{}^{12}{\rm C}$ fusion}

The THM is a powerful and unique indirect technique that allows one to measure the astrophysical factors of the resonant reactions at low energies, where direct methods are not able to obtain data due to very small cross sections. The method was originally suggested by Baur \cite{Baur} but became well known and one of the powerful indirect methods due to  the  leadership of Prof. Claudio Spitaleri \cite{Spitaleri2019,Spitaleri,reviewpaper}. The criticism of the THM in this Letter is not aimed to taint the whole method which demonstrated  its power 
in more than hundred publications. We critically review only the analysis of the data in \cite{THM(2018)}. 

The THM resonant reaction $a+A \to s+F^{*} \to s+ b+B$
where $a=(xA)$ is the the Trojan horse (TH) particle, is described by the two-step mechanism in which the first step is transfer reaction $a+A \to s+F^{*}$  populating the resonance state $F^{*}=x+A$, and the second step is decay of the resonance $F^{*} \to b+B$. Thus, the THM reaction is the process leading to three particles in the final state making analysis of such a reaction quite complicated. Special kinematical conditions should be  fulfilled and angular correlation data are needed to make sure that the reaction is dominantly contributed by the resonant THM mechanism. It allows one to extract from the THM reaction an information about resonant binary sub-reaction $x+A \to F^{*} \to b+B$. 
In \cite{THM(2018)} were measured  the ${}^{12}{\rm C}({}^{14}{\rm N},d)(\alpha + {}^{20}
{\rm Ne})$ and  ${}^{12}{\rm C}({}^{14}{\rm N},d)(p + {}^{24}{\rm Na})$  reactions to obtain the $S^{*}$ factors for the carbon-carbon fusion.
In \cite{THM(2018)} was reported a sharp rise of the astrophysical $S$-factor for carbon-carbon fusion determined using the indirect THM. 
Here are outlined the most important inconsistencies in the THM analysis and data from \cite{THM(2018)}. All the notations are given in \cite{muk2019}. \\

\begin{figure}[htbp]
\includegraphics[width=\columnwidth]{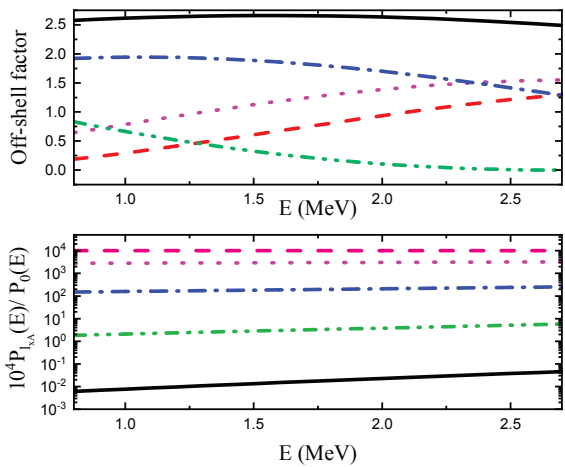}
\caption{(Color online) The upper panel: the off-shell factors $\big|{\cal 
W}_{l_{xA}}\big|^{2}$ for the THM reaction ${}^{12}{\rm C}({}^{14}{\rm N},d){}^{24}{\rm Mg}$ as functions of the ${}^{12}{\rm C}-{}^{12}{\rm C}$
relative energy $E$ at the channel radius $R_{ch}=5$ fm calculated for 5 different ${}^{12}{\rm C}-{}^{12}{\rm C}$ relative orbital angular momenta $l_{xA}$. Red dotted line - $l_{xA}=0$; magenta dotted line- $l_{xA}= 2$; blue dashed-dotted line- $l_{xA}=4$;
green dashed-dotted-dotted line- $l_{xA}=6$ and black solid line - $l_{xA}=8$. The bottom panel: the 
ratio $P_{l_{xA}}/(E)P_{0}(E)$ of the penetrability factors calculated as function of the energy $E$ calculated for the channel radius $R_{ch}= 5$ fm and different $l_{xA}$; $P_{0}$ is the penetrability factor for $l_{xA}=0$. The notations for the lines are the same as in the upper panel. }
\label{fig_PlWl1}
\end{figure}

1. To analyze the measured data Tumino et al. \cite{THM(2018)} used a simple plane-wave approximation (PWA). This approximation neglects the Coulomb interactions between the fragments. The PWA follows from the more general distorted-wave-Born approximation (DWBA) in which the distorted waves are replaced by the plane waves and can be used only if the PWA calculations provide a reasonable agreement with DWBA ones. It is not the case under consideration \cite{muk2019}.
It has been demonstrated in \cite{muk2019}  that the rise of the $S^{*}$ factors at low energies seen in the aforementioned work was an artifact of using the PWA, which is not usable for the case under consideration. It was shown that such a rise disappears if the Coulomb (or Coulomb-nuclear) interactions in the initial and  final states are included.
To support the statement that the PWA cannot be used for the analysis of the THM reaction under consideration, we present the excerpt from the first work where the reaction 
${}^{12}{\rm C}({}^{14}{\rm N},d){}^{24}{\rm Mg}$ was discussed \cite{Nagatani}: ".. the DWBA should provide a straightforward means to describe the dynamics of the reaction.."  Another very compelling evidence  that the PWA should not be used is presented in the pioneering work \cite{Goldberg} where the experimental angular distribution of the deuterons from the THM reaction ${}^{12}{\rm C}({}^{14}{\rm N},d){}^{24}{\rm Mg}$ at $33$ MeV incident energy of ${}^{14}{\rm N}$. In this experiment the incident energy was higher than in the THM experiment in \cite{THM(2018)}. Besides the excited bound state of ${}^{24}{\rm Mg}$ was populated rather than the resonance state.  All of these what makes the energy of the  outgoing deuterons higher than the Coulomb barrier in the final  $d+{}^{24}{\rm Mg}$ state (note that in the THM experiment the energies of deuterons were below the Coulomb barrier).  Nevertheless, the experimental angular distribution was flat and was perfectly reproduced by the DWBA calculation,  which agrees with the DWBA calculation in \cite{muk2019}. Moreover, it follows from \cite{Goldberg} that the momentum distribution of the deuterons disagrees with the one extracted in \cite{THM(2018)}. It casts doubt about the mechanism measured in \cite{THM(2018)}.  Note that the deuteron angular distribution was not presented  in \cite{THM(2018)}. The only criterium  used in the THM analysis to justify the PWA  was the deuteron momentum distribution.  It worked for lighter nuclei \cite{Spitaleri2019}  but not in the case under consideration in which different competing mechanisms do contribute, such as ${}^{10}{\rm B}$ transfer from the ${}^{12}{\rm C}$ target to ${}^{14}{\rm N}$ or  ${}^{8}{\rm Be}$ transfer to ${}^{12}{\rm C}$ leaving ${}^{6}{\rm Li}$ in the resonance state decaying into $d+\alpha$ channel. 
  The angular correlations of the final-state particles, which provide the most crucial information needed to identify 
  the reaction mechanism \cite{Goldberg}, are missing in \cite{THM(2018)}.\\
2. The THM double differential cross section (DCS)  in \cite{THM(2018)}  does not correspond to the one described by the two-step THM mechanism for the THM reaction ${}^{14}{\rm N} + {}^{12}{\rm C} \to d+ {}^{24}{\rm Mg}^{*} \to \alpha (p)+{}^{20}{\rm Ne} ({}^{23}{\rm Na})$. The double DCS for the THM mechanism is given by Eq. (39) from \cite{muk2019}. Specifically, for the reaction under consideration described by the THM mechanism, the energy of the outgoing deuterons corresponding to the  ${}^{24}{\rm Mg}$ resonance energy at $E=2.1$ MeV is $E_{d{}^{24}{\rm Mg}}=1.47$ MeV while for $E=2.6$ MeV $E_{d{}^{24}{\rm Mg}}=0.97$ MeV. Thus the deuterons are well below the Coulomb barrier of $3$ MeV in the system  $d + {}^{24}{\rm Mg}$. 
Hence, the DWBA DCS $\frac{{d\sigma }^{DWZR(prior)}}{{d{\Omega _{{{\rm {\bf k}}_{sF}}}}}}$ of the transfer reaction ${}^{14}{\rm N} + {}^{12}{\rm C} \to d+ {}^{24}{\rm Mg}^{*}$, which is the first step of the THM reaction, drops by two orders of magnitude on the interval $2.1 - 2.64$ MeV interval what should lead to the decrease of the THM double DCS.  Hence, one can expect that as energy $E$ increases the non-THM mechanisms, which are background, should dominate. These mechanisms were not identified (see the discussion above).
 That is why we suspect that the THM  $S^{*}$ factor at $E> 2.2$ MeV was contributed by the background (see Fig.  \ref{fig3Tang}). \\
3.  Now we address to the comparison of the THM $S$ factors with the experimental ones. This discussion concerns a general matter of the application of the THM but is especially important for the application of the THM for heavier nuclei like the case under consideration. The main advantage of the THM is the absence of the Coulomb-centrifugal barrier in the entry channel of the 
${}^{12}{\rm C}-{}^{12}{\rm C}$. That is why in the THM one can observe both low and high $l_{xA}$ resonances.
    There is the price one pays for the absence of the barrier: the transferred particle 
    $x={}^{12}{\rm C}$ in the THM reaction is off-the-energy shell. As the result in the double DCS appears the off-shell factor $\big|{\cal W}_{l_{xA}}\big|^{2}$, see Eqs. (39) and  (32) from \cite{muk2019}.
    This off-shell factor can increase the contribution from high spin resonances, which are suppressed in direct measurements.
    In Fig. \ref{fig_PlWl1} is shown the off-shell factor $\big|{\cal W}_{l_{xA}}\big|^{2}$ 
    calculated for the ${}^{12}{\rm C}({}^{14}{\rm N},d){}^{24}{\rm Mg}$ reaction at the incident energy of $\,{}^{14}{\rm N}\,$ $\,30\,$ MeV and different $l_{xA}$. 
The THM brings two modifications to the DCS of the binary resonant sub-reactions $x+ A \to F^{*} \to b+B$. It removes the penetrability factor $P_{l_{xA}}$ in the entry channel of the sub-reaction and each partial wave is multiplied by the off-shell factor $\big|{\cal 
W}_{l_{xA}}\big|^{2}$, which may significantly modify 
the relative weight of the resonances with different $l_{xA}$. The selected channel radius $R_{ch}=5$ fm corresponds to the grazing collision of two carbon nuclei.  From the bottom panel of Fig. \ref{fig_PlWl1}  one can conclude that the dominant contribution to the carbon-carbon fusion in direct measurements at low energies comes from two partial waves: $l_{xA}=0$ and $2$. It is also confirmed by calculations in \cite{ErbBromley}. 
However, in the THM the dominant contribution comes from high spins.  This effect is called kinematical enhancement of higher spin resonances. In particular, in \cite{THM(2018)}  
the resonance spins up to $6$ were assigned at energies below $1.6$ MeV. The assigned in \cite{THM(2018)} spin $1^{-}$ for the resonance at $0.877$ MeV is a mistake because a resonance with negative parity cannot be populated in the collision of two identical bosons such as ${}^{12}{\rm C}$ nuclei.  The difference between the low-energy resonance spins contributed to the direct and indirect THM measurements should be taken into account  when comparing the direct and THM $S^{*}$ factors.\\
4. Despite of the criticism of the analysis in \cite{THM(2018)} the experimental THM double DCS reveals two strong resonances at $0.88$ and $1.5$ MeV what is an undeniable achievement of the THM because such low energies are not yet reachable in direct measurements. A strong resonance at $1.5$ MeV, which is inside the Gamow window, is of a crucial importance for the ${}^{12}{\rm C}-{}^{12}{\rm C}$ fusion rates  and will play the same role as the "Hoyle" state \cite{Hoyle54} in the triple-$\alpha$ process of synthesis of ${}^{12}{\rm C}$  \cite{Freer14}. 

Note that the models using the global potentials in \cite{Esbensen,Wiescher,Rowley,Assuncao} can predict resonances only above $E = 2.5$ MeV (the TDHF calculations of \cite{Godbey} do not take into account any possible resonant behavior in the fusion process). The THM data at low energies call for an improvement of the theoretical models.

There have been predictions based on phenomenological considerations of explosive stellar events such as superbursts that suggest a  strong $^{12}$C+$^{12}$C cluster resonance around $E = 1.5$ MeV in $^{24}$Mg that would drastically enhance the energy production and may provide a direct nuclear driver for the superburst phenomenon. However, no indication for such a state has not yet been reported in direct measurements in which the minimal measured energy is $E \approx 2.1$ MeV. Hence, the discovery of two strong resonances at $0.88$ and $1.5$ MeV in the THM data is an important contribution of this method in the ${}^{12}{\rm C}-{}^{12}{\rm C}$ fusion research.

\section{Conclusion}
A comparison of the direct measurements of ${}^{12}{\rm C}- {}^{12}{\rm C}$
fusion and the indirect THM data \cite{THM(2018)} is presented. We found disagreements between the direct and THM $S$ factors. We also outlined shortcomings in the THM analysis. Moreover, the very recent direct data by Notre Dame University \cite{Wiescher2019} show even more stronger disagreement between the direct and  the THM $S$ factors from \cite{THM(2018)}. Taking into account the critical importance of the 
${}^{12}{\rm C}+{}^{12}{\rm C}$ fusion in many astrophysical scenarios we can outline three major problems to be solved in the near future:\\
1. Extending direct measurements below $2.1$ MeV with the final goal to reach $1.5$ MeV. It would help us to better understand the trend of the $S^{*}$ factors at energies $E < 2.1$ MeV and resolve the question about the validity of the hindrance model. \\
2. Perform new indirect measurements. One such preparation is underway at the Cyclotron Institute, Texas A\&M University \cite{Rogachev}.\\
3. Improving of the theoretical models, which will be able to predict the low-energy resonances detected in the THM.  

\acknowledgments{A.M.M. acknowledges a support from the U.S. DOE Grant No. DE-FG02-93ER40773 and the NNSA Grant No. DENA000384. 
X.T. is supported in part by the National Key Research and Development program (MOST 2016YFA0400501) from the Ministry of Science and Technology of China and from the key research program (XDPB09-2).
A.M.M. thanks Dr. Goldberg for very useful comments.}

\begin{center}

\end{center}
\end{document}